\documentclass[epj,final]{svjour}
\usepackage{graphics}
\begin{document}
\title{Doubly heavy quark baryon spectroscopy and semileptonic decay}
\author{C. Albertus\inst{1} \and E. Hern\'andez\inst{2}
\and J. Nieves \inst{1} \and J.M. Verde-Velasco \inst{2}
}                     % Do not remove
\institute{ Departamento de F\'{\i}sica At\'omica, Molecular y
  Nuclear, Universidad de Granada, E-18071 Granada, Spain\and Grupo de
  F\'{\i}sica Nuclear, Departamento de F\'{\i}sica Fundamental e
  IUFFyM, Facultad de Ciencias, E-37008 Salamanca, Spain.}
\date{Received: date / Revised version: date}
\abstract{
Working in the framework  of a
nonrelativistic quark model we evaluate the spectra and semileptonic
decay widths for the ground state
of doubly heavy $\Xi$ and $\Omega$ baryons.  We solve the three-body problem using
a variational ansatz  made possible by the constraints imposed
by heavy quark spin symmetry. In order to check the dependence of our results
on the inter-quark interaction we have used five different quark
quark potentials which include Coulomb and hyperfine terms coming from
one-gluon exchange, plus a confining term. Our results for the spectra are in
good agreement with a previous calculation done using a Faddeev approach.
For the semileptonic decay  our results for the total decay widths are in a good
agreement with the ones obtained within a relativistic quark
model in the quark-diquark approximation.
 % end of PACS codes
} %end of abstract
\maketitle
\section{Introduction}
Even though only recently the mass of a baryon with two heavy quarks
has been measured experimentally\cite{mattson02}, these systems have been 
being studied for more than a decade. Working with  a system with two heavy quarks 
one can take advantage of the constraints imposed by heavy quark spin symmetry
(HQSS).
This symmetry amounts to the decoupling of the heavy quark spins in the
infinity heavy quark mass limit. In that limit one can consider the total spin
 of the two heavy quarks subsystem to be well defined. This result, that we
 shall assume to be  valid for the actual heavy quark masses,
  will simplify the solution of the three-body problem. 

In this contribution  we shall present results for masses and total semileptonic
decay widths. We have also analyzed other static observables  as well as form
factors, differential decay widths and angular asymmetries of the weak decays.
For a detailed account of  the full calculation see Ref. \cite{paper}
 
 In Table \ref{tab:summ} we summarize the quantum numbers
of the baryons considered in this study.

\begin{table}

\caption{Quantum numbers of doubly heavy baryons analyzed in this study. $S$, $J^P$ are strangeness and the spin parity
of the baryon, $I$ is the isospin, and
$S_{h}^\pi$ is the spin parity of the heavy
degrees of freedom. $l$ denotes a light  $u$ or
$d$ quark .}

\begin{tabular}{cccccccc}\hline
Baryon &~~~$S$~~~&~~~$J^P$~~~&~~~ $I$~~~&~~~$S_{ h}^\pi$~~~& 
Quark 

\\
       &       &         &   &          &  content               
\\\hline
$\Xi_{cc}$& 0 &$\frac12^+$& $\frac12$ &$1^+$&$ccl$
\\
$\Xi^*_{cc}$ & 0 &$\frac32^+$&$\frac12$  &$1^+$&$ccl$
\\
$\Omega_{cc}$ & $-1$ &$\frac12^+$& 0 &$1^+$&$ccs$
\\
$\Omega^*_{cc}$ & $-1$ &$\frac32^+$&$0$&$1^+$&$ccs$
\\\\
$\Xi_{bb}$& 0 &$\frac12^+$& $\frac12$ &$1^+$&$bbl$
\\
$\Xi^*_{bb}$ & 0 &$\frac32^+$&$\frac12$  &$1^+$&$bbl$
\\
$\Omega_{bb}$ & $-1$ &$\frac12^+$& 0 &$1^+$&$bbs$
\\
$\Omega^*_{bb}$ & $-1$ &$\frac32^+$&$0$&$1^+$&$bbs$
\\\\
$\Xi'_{bc}$& 0 &$\frac12^+$& $\frac12$ &$0^+$&$bcl$
\\
$\Xi_{bc}$ & 0 &$\frac12^+$&$\frac12$  &$1^+$&$bcl$
\\
$\Xi^*_{bc}$& 0 &$\frac32^+$& $\frac12$ &$1^+$&$bcl$
\\
$\Omega'_{bc}$ & $-1$ &$\frac12^+$& 0 &$0^+$&$bcs$
\\
$\Omega_{bc}$ & $-1$ &$\frac12^+$& 0 &$1^+$&$bcs$
\\
$\Omega^*_{bc}$ & $-1$ &$\frac32^+$&$0$&$1^+$&$bcs$
\\
\hline
\end{tabular}

\label{tab:summ}
\end{table}
   
\section{The model}

Once the centre of mass (CM) motion has been removed, the intrinsic
Hamiltonian that describes the inner dynamics of the baryon is given
by 
\begin{eqnarray}
 H^{\rm
int}&=&\sum_{j=1,2}\,H_j^{sp}+V_{h_1h_2}(\vec r_1-\vec r_2,\, spin)-
\frac{\vec{\nabla}_1\cdot\vec{\nabla}_2}
{m_q}+\overline M
\nonumber
\\
H_j^{sp}&=&-\frac{\vec{\nabla}\stackrel{}{^2}_{\hspace{-0.2cm}j}
}{2\mu_j}+V_{h_jq}(\vec r_j,\, spin),\ j=1,2
\end{eqnarray}
where  $\vec{r}_1,\, \vec{r}_2$ are the relative positions
of the $h_1,\,h_2$ heavy quarks with respect to the light quark $q$, 
 $\overline M = m_{h_1}+m_{h_2}+m_{q}$, $\mu_{j} = \left (
1/m_{h_j} + 1/m_q\right)^{-1}$ and $\vec{\nabla}_{j} =
\partial/\partial_{\vec{r}_j},\ j=1,2$.~$V_{h_jq}$ and $V_{h_1h_2}$ are
the heavy-light and heavy-heavy interaction potentials. Note 
 the presence of the Hughes-Eckart term that results from the
separation of  the CM  motion.

For the  quark quark interaction we have considered five different
phenomenological potentials, one suggested by Bhaduri and
collaborators ~\cite{BD81} (BD) and four suggested by
B. Silvestre-Brac and C. Semay~\cite{Si96,SS93} (AL1, AL2, AP1 y
AP2). All of them  include Coulomb and hyperfine terms coming from
one-gluon exchange and a confining term, and differ in the form factor
used for the hyperfine term, the use of a form factor in the one gluon
exchange Coulomb term or in the power of the confinement term. All free
parameters had been adjusted to reproduce the light and heavy-light
meson spectra. Details  on the potentials can be found in Refs.~  
\cite{BD81,Si96,SS93}.

For the interactions considered, the total spin and internal orbital
angular momentum commute with the intrinsic Hamiltonian, and thus are
well defined. In this work we will study the ground state of baryons
with total angular momentum $J=1/2,~ 3/2$ so we can assume the orbital
angular momentum to be $0$. This implies that the spatial wave
function only depend on  $r_1$, $r_2$ and
$r_{12}=|\vec{r}_1-\vec{r}_2|$. We will also assume that taking the
total spin of the heavy degrees of freedom to be well defined, as obtained in
the infinite heavy quark mass limit, is a good approximation. That will
allow us to  write the wave function in a simple way (see Ref.
\cite{paper} for details).

The spatial part of wave function will be determined using a variational
method in which we will assume the following  functional form: 
\begin{eqnarray}
\Psi_{h_1h_2}^{B} (r_1,r_2,r_{12}) &=& N\, F^{B}(r_{12})\,
\phi_{h_1q}(r_1)\,\phi_{h_2q}(r_2)
\end{eqnarray}
where  $N$ is a normalization constant, $\phi_{h_jq}$ is the $S$-Wave
ground state wave function  $\varphi_j(r_j)$ of the single particle
Hamiltonian $H_j^{sp}$ corrected at large distances:
\begin{eqnarray}
\phi_{h_jq} (r_j) &=& (1+\alpha_j\,r_j)\,\varphi_j(r_j),\ \ j=1,2
\end{eqnarray}
The heavy-heavy Jastrow correlation function $F^B$ will be given as a
linear combination of gaussians: 
\begin{eqnarray}
F^{B}(r_{12}) &=& 
\sum_{j=1}^4 a_j e^{-b_j^2(r_{12}+d_j)^2},\quad a_1=1 
\label{eq:f12}
\end{eqnarray}
where  $\alpha_i$, $a_i~ i\ne 1$, $b_i$ and $d_i$ are free variational
parameters. The values that we get for the variational parameters  
are compiled in  Ref. \cite{paper}.

We have also used the wave function obtained in this model to study
different doubly $B(1/2^+)\to B'(1/2^+)$ baryon semileptonic decays involving a $b\to c $
transition at the quark level. We have worked in the spectator
approximation with only one-body currents. 
\\
The differential decay width reads
\begin{eqnarray}
&&{\rm d}\Gamma= 
8 |V_{cb}|^2 m_{B'} G_F^{\,2}  
 \frac{d^3p^\prime}{(2\pi)^32E^\prime_{B'} }
\frac{d^3k}{(2\pi)^32E_{\bar \nu_l} } \frac{d^3k^\prime}{(2\pi)^32E^\prime_{l}
}  (2\pi)^4  
\nonumber
\\
&&\hspace{1cm}\times\delta^4(p-p^\prime-k-k^\prime)\ {\cal
L}^{\alpha\beta}(k,k')
{\cal H}_{\alpha\beta}(p,p')    
\end{eqnarray}
where $|V_{cb}|$ is the modulus of the corresponding Cabibbo--Kobayashi--Maskawa
matrix element, $m_{B'}$ is the mass of the final baryon, 
$G_F$ is the Fermi decay constant, $p$, $p'$,
$k$ and $k'$ are the four-momenta of the initial baryon, final baryon, final
anti-neutrino and final lepton respectively,  and 
${\cal L}$  and ${\cal H}$ are the lepton and hadron tensors.\\ 
The lepton tensor
is given as
\begin{eqnarray}
{\cal L}^{\mu\sigma}(k,k')&=& k'^\mu k^\sigma +k'^\sigma k^\mu
- g^{\mu\sigma} k\cdot k^\prime + {\rm i}
\epsilon^{\mu\sigma\alpha\beta}k'_{\alpha}k_\beta \label{eq:lep}
\end{eqnarray}
where we use the convention $\epsilon^{0123}=-1$,
$g^{\mu\mu}=(+,-,-,-)$. \\
The hadron tensor is given as
\begin{eqnarray}
&& {\cal H}_{\mu\sigma}(p,p') = \nonumber
\\ && \hspace{.4cm}
\frac12 \sum_{r,r'}  
 \left\langle B', r'\
\vec{p}^{\,\prime}\left|\,
\overline \Psi^c(0)\gamma_\mu(I-\gamma_5)\Psi^b(0)\right| B, r\ \vec{p}   \right\rangle 
\nonumber \\
&& \hspace{1cm} \times
\ \left\langle B', r'\ 
\vec{p}^{\,\prime}\left|\,\overline \Psi^c(0)\gamma_\sigma(I-\gamma_5)
\Psi^b(0) \right|  B, r\ \vec{p} \right\rangle^*
\label{eq:wmunu}
\end{eqnarray}
with $\left|B, r\ \vec p\right\rangle\, (\left|B', r'\ \vec{p}\,'\right\rangle)$
representing the initial (final) baryon with three--momentum $\vec p$
($\vec{p}\,'$)
and spin index $r$ ($r'$). The baryon states are normalized such that
\begin{equation}
\langle r\ \vec{p}\, |\, r'
\ \vec{p}^{\,\prime} \rangle
= (2\pi)^3 (E(\vec p\,)/m)\,\delta_{rr'}\,
\delta^3(\vec{p}-\vec{p}^{\,\prime})
\end{equation}

\noindent
We compute the widths  similarly
as we did in Ref. \cite{semilep} for  baryons with a heavy quark.

\section{Results and discussion}
%---------------------------------------
\begin{table}
\caption{Doubly heavy $\Xi$ masses in MeV.}
\label{masaXi}
\begin{tabular}{lcccc}
& This work&\cite{Si96}&
Exp.~\cite{mattson02}
&Lattice~\cite{flynn03}\\\hline \\
$\Xi_{cc}$ \hspace{0.5cm} &$3612^{+17}$ &$3609^{+22}$&$3519\pm1$&$3549\pm 95$ \\   
$\Xi^*_{cc}$  &$3706^{+23}$ &        &  &$3641\pm97$\\   
$\Xi_{bb}$  &$10197^{+10}_{-17}$&$10194^{+10}_{-19}$ &\\   
$\Xi^*_{bb}$  &$10236^{+9}_{-17}$&\\   
$\Xi_{bc}$  &$6919^{+17}_{-7}$ &$6916^{+18}_{-9}$  \\   
$\Xi'_{bc}$  &$6948^{+17}_{-6}$ \\   
$\Xi^*_{bc}$  &$6986^{+14}_{-5}$\\ \hline \\

\end{tabular}
%\end{table}

%\begin{table}[h]
%\label{tab:xiM2}
%\caption{caption}
\begin{tabular}{lccccccc}
& \hspace{-.3cm}This work &\hspace{-.21cm}
\cite{ebert02}&\hspace{-.21cm}\cite{kiselev02}&\hspace{-.21cm}\cite{tong00}&\hspace{-.21cm}\cite{gershtein00}&\hspace{-.21cm}\cite{roncaglia95}&\hspace{-.21cm}\cite {Vijande:2004at} \\\hline \\
$\Xi_{cc}$ &\hspace{-.3cm}    $3612^{+17}$
&\hspace{-.21cm}3620&\hspace{-.21cm}3480&\hspace{-.21cm}3740&\hspace{-.21cm}
3478&\hspace{-.21cm}3660 &\hspace{-.21cm} 3524\\   
$\Xi^*_{cc}$  &\hspace{-.3cm} $3706^{+23}$ &\hspace{-.21cm}3727&\hspace{-.21cm}3610&\hspace{-.21cm}3860&\hspace{-.21cm}3610&\hspace{-.21cm}3740&\hspace{-.21cm} 3548\\   
$\Xi_{bb}$ & \hspace{-.3cm}   $10197^{+10}_{-17}$ &\hspace{-.21cm}10202&\hspace{-.21cm}10090&\hspace{-.21cm}10300&\hspace{-.21cm}10093&\hspace{-.21cm}10340\\   
$\Xi^*_{bb}$& \hspace{-.3cm}  $10236^{+9}_{-17}$ &\hspace{-.21cm}10237&\hspace{-.21cm}10130&\hspace{-.21cm}10340&\hspace{-.21cm}10133&\hspace{-.21cm}10370\\   
$\Xi_{bc}$ & \hspace{-.3cm}   $6919^{+17}_{-7}$ &\hspace{-.21cm}6933&\hspace{-.21cm}6820&\hspace{-.21cm}7010&\hspace{-.21cm}6820&\hspace{-.21cm}7040 \\   
$\Xi'_{bc}$ & \hspace{-.3cm}  $6948^{+17}_{-6}$ &\hspace{-.21cm}6963&\hspace{-.21cm}6850&\hspace{-.21cm}7070&\hspace{-.21cm}6850&\hspace{-.21cm}6990\\   
$\Xi^*_{bc}$ & \hspace{-.3cm} $6986^{+14}_{-5}$ &\hspace{-.21cm}6980&\hspace{-.21cm}6900&\hspace{-.21cm}7100&\hspace{-.21cm}6900&\hspace{-.21cm}7060 \\   
\hline
\end{tabular}

\end{table}

%--------------------------

%---------------------------------------------------------------------------------
%---------------------------------------------------------------------------------

\begin{table}
\caption{Doubly heavy $\Omega$ masses in MeV.}
\label{tab:OmM1}
\begin{tabular}{lccc}
& This work&\cite{Si96}
&Lattice~\cite{flynn03}\\\hline\\
$\Omega_{cc}$\hspace{.5cm}  &$3702^{+41}$  &$3711^{+30}_{-2}$& $3663\pm 97$\\   
$\Omega^*_{cc}$ &$3783^{+22}$ && $3734\pm 98$ \\   
$\Omega_{bb}$  &$10260^{+14}_{-34}$\\   
$\Omega^*_{bb}$ &$10297^{+5}_{-28}$ \\   
$\Omega_{bc}$   &$6986^{+27}_{-17}$ &$7003^{+20}_{-32}$&\\   
$\Omega'_{bc}$  &$7009^{+24}_{-15}$\\   
$\Omega^*_{bc}$ &$7046^{+11}_{-9}$\\ \hline\\  
\end{tabular}

%\end{table}
%
%\begin{table}[h]
%\caption{Caption}
%\label{tab:OmM2}
\begin{tabular}{lcccccc}
& This work&
\cite{ebert02}&\cite{kiselev02}&\cite{tong00}
&\cite{gershtein00}&\cite{roncaglia95}\\\hline\\
$\Omega_{cc}$  &$3702^{+41}$  & 3778&3590&3760&3590&3740\\   
$\Omega^*_{cc}$ &$3783^{+22}$ &  3872&3690&3900&3690&3820 \\   
$\Omega_{bb}$  &$10260^{+14}_{-34}$ &
10359&10180&10340&10180&10370\\   
$\Omega^*_{bb}$ &$10297^{+5}_{-28}$ & 10389&10200&10380&10200&10400\\   
$\Omega_{bc}$   &$6986^{+27}_{-17}$ &7088&6910&7050&6910&7090\\   
$\Omega'_{bc}$  &$7009^{+24}_{-15}$ &7116&6930&7110&6930&7060\\   
$\Omega^*_{bc}$ &$7046^{+11}_{-9}$ &7130&6990&7130&6990&7120 \\   
\hline
\end{tabular}

\end{table}

\noindent
The mass of the baryon is simply given by the expectation value of the
intrinsic Hamiltonian.
In table \ref{masaXi}  we give our results for doubly heavy $\Xi$
baryons, while in table \ref{tab:OmM1} are the results for the
doubly heavy $\Omega$ ones. Our central values correspond to the results obtained using
the AL1 potential, while the errors quoted take into account the
variations found when using the other potentials. That also applies to
the quoted results for Ref. \cite{Si96}, obtained with the same
interaction potentials but within a Faddeev approach. When comparison with this
work is possible we find an excellent agreement between the two
calculations. Besides we give predictions for states not considered in
the study of Ref. \cite{Si96}. We also compare  with other
theoretical models. All calculations give similar results that
vary within a few percent. 
From the experimental side  only the mass of the $\Xi_{cc}$ has been measured.
The experimental value for $M_{\Xi_{cc}}$ obtained
by the SELEX Collaboration~\cite{mattson02} is $100$ MeV
smaller than our result. Note nevertheless that no account is given 
of the systematic error. There are also
 lattice calculations, by the
UKQCD Collaboration \cite{flynn03}, of the masses of the doubly charmed
 $\Xi_{cc},\,\Xi^*_{cc},\,\Omega_{cc}$ and $\Omega^*_{cc}$ baryons. 
 Our results are within errors of the lattice determinations.

%-----------------------------------------------------
%----------------------------------------------------
\begin{table}
\caption{Semileptonic decay widths in units of $10^{-14}$\,GeV. We have
used  $|V_{cb}|=0.0413$. $l$ stands for $l=e,\,\mu$}
\begin{tabular}{lccccc}
  &This work &\cite{ebert04}&\cite{faessler01}&\cite{guo98}&\cite{sanchis95}\\ \hline\\
$\Gamma(\Xi_{bb}\to\Xi_{bc}\,l\bar \nu_l)$\hspace{.25cm}  &$ 3.84^{+0.49}_{-0.10}$&
3.26&&28.5&\\ \\
$\Gamma(\Xi_{bc}\to\Xi_{cc}\,l\bar \nu_l)$  &$ 5.13^{+0.51}_{-0.05} $ & 4.59&0.79&8.93&4.0\\ \\
$\Gamma(\Xi_{bb}\to\Xi_{bc}'\,l\bar \nu_l)$ &  $2.12 ^{+0.26}_{-0.05}$ &1.64&&4.28&\\ \\
$\Gamma(\Xi_{bc}'\to\Xi_{cc}\,l\bar \nu_l)$ &  $2.71^{+0.19}_{-0.05}$ & 1.76&&7.76&\\ \hline
\end{tabular}
%\end{table}
%
%\begin{table}
\begin{tabular}{lccc}
  &This work &\cite{ebert04}&\cite{guo98}\\ \hline\\
$\Gamma(\Omega_{bb}\to\Omega_{bc}\,l\bar \nu_l)$\hspace{.25cm}  & $4.28^{+0.39}_{-0.03}$& 3.40&28.8\\ \\
$\Gamma(\Omega_{bc}\to\Omega_{cc}\,l\bar \nu_l)$ &$5.17^{+0.39}$&4.95&\\ \\
$\Gamma(\Omega_{bb}\to\Omega_{bc}'\,l\bar \nu_l)$ &$2.32^{+0.26}$& 1.66&\\ \\
$\Gamma(\Omega_{bc}'\to\Omega_{cc}\,l\bar \nu_l)$ &$2.71^{+0.17} $ &1.90&\\\hline
\end{tabular}

\label{tab:gamma}
\end{table}%

In table \ref{tab:gamma} we present our results for the semileptonic
decay widths for the different processes under study. Our central
values again correspond to the results obtained using the AL1
potential, while the errors show the variations when using the other
four potentials. The biggest variations appear for the BD potential, with
differences of the order of \mbox{$7 \sim 12 \%$}. We compare our
results with the predictions of different models. For that purpose we
need to fix a value for $|V_{cb}|$ for which we take
$|V_{cb}|=0.0413$.  Our results are
 in reasonable agreement with the ones  in Ref.~\cite{ebert04} where
 they
 use a
 relativistic quark model evaluated in the quark-diquark
approximation. For $\Gamma(\Xi_{bc}\to\Xi_{cc})$
 we also agree with the value of Ref.~\cite{sanchis95} obtained using heavy 
 quark effective theory. A much smaller value for the same width 
 is obtained in   the relativistic   three--quark model calculation of
Ref.~\cite{faessler01}.
In Ref.~\cite{guo98}, where they use
 the Bethe--Salpeter equation applied to a quark-diquark
 system, they obtain  much larger
  results for all transitions.

%%
%\section{Introduction}
%\label{intro}
%Your text comes here. Separate text sections with
%\section{Section title}
%\label{sec:1}
%and \cite{RefJ}
%\subsection{Subsection title}
%\label{sec:2}
%as required. Don't forget to give each section
%and subsection a unique label (see Sect.~\ref{sec:1}).
%%
%% For one-column wide figures use
%\begin{figure}
%% Use the relevant command for your figure-insertion program
%% to insert the figure file.
%% For example, with the option graphics use
%\resizebox{0.75\textwidth}{!}{%
%  \includegraphics{leer.eps}
%}
%% If not, use
%%\vspace{5cm}       % Give the correct figure height in cm
%\end{figure}
%%
%% For two-column wide figures use
%\begin{figure*}
%% Use the relevant command for your figure-insertion program
%% to insert the figure file. See example above.
%% If not, use
%\vspace*{5cm}       % Give the correct figure height in cm
%\end{figure*}
%\begin{table}
%\begin{tabular}{lll}
%\hline\noalign{\smallskip}
%\end{tabular}
%% Or use
%\vspace*{5cm}  % with the correct table height
%\end{table}
%%

\begin{acknowledgement}
 This research was supported by DGI and FEDER funds, under contracts
FIS2005-00810, BFM2003-00856 and FPA2004-05616, by Junta de
Andaluc\'\i a and Junta de Castilla y Le\'on under contracts FQM0225
and SA104/04, and it is part of the EU integrated infrastructure
initiative Hadron Physics Project under contract number
RII3-CT-2004-506078. C. A. acknowledges a research contract with Universidad de
Granada. J. M. V.-V. acknowledges an  E.P.I.F. contract with
Universidad de Salamanca.
\end{acknowledgement}\vspace*{-.5cm}

% BibTeX users please use
% \bibliographystyle{}
% \bibliography{}

\end{document}